\documentclass[aps,showpacs,showkeys,preprint]{revtex4}
\usepackage[dvips]{graphicx}
\usepackage{latexsym,amsmath,amssymb}

\newcommand{\reprd}[3]{Phys.\ Rev.\ D \textbf{#1} (#2) #3}
\newcommand{\reprl}[3]{Phys.\ Rev.\ lett.\ \textbf{#1} (#2) #3}
\newcommand{\pr}[3]{Phys.\ Rev.\ \textbf{#1} (#2) #3}
\newcommand{\cmp}[3]{Commun.\ Math.\ Phys.\ \textbf{#1} (#2) #3}

\newcommand{\cqg}[3]{Class.\ Quant.\ Grav.\ \textbf{#1} (#2) #3}

\newcommand{\grg}[3]{Gen.\ Rel.\ Grav.\ \textbf{#1} (#2) #3}
\newcommand{\ijmpa}[3]{Int.\ J.\ Mod.\ Phys.\ A \textbf{#1} (#2) #3}
\newcommand{\ijtp}[3]{Int.\ J.\ Theor.\ Phys.\ \textbf{#1} (#2) #3}

\newcommand{\lnc}[3]{Lett.\ Nuovo.\ Cim.\ \textbf{#1} (#2) #3}

\newcommand{\plb}[3]{Phys.\ lett.\ B \textbf{#1} (#2) #3}

\newcommand{\hepth}[1]{hep-th/#1}
\newcommand{\grqc}[1]{gr-qc/#1}
\newcommand{\arXivid}[1]{arXiv:#1}

\begin{document}


 \title{Thermodynamics of (2+1)-dimensional acoustic black hole
 based on the generalized uncertainty principle}


\author{Wontae Kim$^{a,b,c}$}
  \email{wtkim@sogang.ac.kr}
\author{Edwin J. Son$^{a,c}$}
  \email{eddy@sogang.ac.kr}
\author{Myungseok Yoon$^b$}
  \email{younms@sogang.ac.kr}
  \affiliation{$^a$Department of Physics, Sogang University, Seoul
    121-742, Korea\\ 
    $^b$Center for Quantum Spacetime, Sogang University, Seoul
    121-742, Korea\\ 
    $^c$Basic Science Research Institute, Sogang University, Seoul
    121-742, Korea} 

\date{\today}

\begin{abstract}
We study thermodynamic quantities of an acoustic black hole and its
thermodynamic stability in a cavity based on the generalized uncertainty principle.
It can be shown that 
there is a minimal black hole which can be a stable remnant 
after black hole evaporation. Moreover, the behavior of the free energy shows
that the large black hole is stable too. Therefore, the acoustic black hole
can decay into the remnant or the large black hole.
\end{abstract}


\maketitle


Bekenstein has suggested that a black hole can have an entropy
proportional to the surface area at the horizon~\cite{bekenstein} and then
Hawking has shown that the Schwarzschild black hole has thermal
radiation with a temperature $T_H = (8\pi M)^{-1}$
by applying the quantum field theory where $M$ is the mass of
the black hole~\cite{hawking}. 
Since then, the thermodynamics has been studied in
various black holes and thermodynamic local quantities have been
calculated in a cavity with a finite size~\cite{gh,hi,bcm,allen,york,wy,brown}. 
Based on the quantum gravity and the string theory, the generalized
uncertainty principle(GUP) with a minimal length~\cite{pad,maggiore,garay,kmm} instead of 
the conventional Heisenberg uncertainty principle (HUP)
has been intensively studied and many interesting applications have been done;
especially, some corrections
to the entropy by the GUP have been 
studied in Ref.~\cite{kln,mv}. Moreover, the GUP has been also applied to study
the thermodynamics and the stability in the Schwarzschild black hole~\cite{acs,ch,mkp,ksy}.
On the other hand, intriguing 
black-hole issues have been well appreciated in the
field theoretical framework of fluid
because the acoustic analogue is useful in that its thermodynamics,
such as the Hawking radiation and the black hole entropy might be tested hopefully in
the laboratory~\cite{unruh}. Indeed, a ``draining bathtub'' referred to as an
acoustic analogue of a rotating black hole has been extensively
studied~\cite{visser,kko,vw,ls,bm,bcl}. 

From the thermodynamic point of view, a small black hole can be
created through the Gross-Perry-Yaffe(GPY) phase transition 
when the temperature is
over the critical one~\cite{gpy,hp,sh} and it
decays into hot flat space, since it is unstable while a large black
hole is stable. In this work, we would like to study thermodynamic quantities
and the stability of the (2+1)-dimensional acoustic black hole 
based on the GUP. 
In the GUP side, there should be a minimal
black hole whose size is comparable to
the minimal length so that it cannot evaporate completely through the thermal
radiation. It follows from the careful consideration of 
the local temperature and the heat capacity on the 
boundary of the cavity with a finite size.
For this purpose, we shall introduce the acoustic black hole and
study its thermodynamics defined in the conventional HUP 
in the cavity, and then extend it to the GUP regime.
It can be shown that 
there is a minimal black hole which can be a
stable remnant. So, the acoustic black hole
can decay into a stable remnant or a stable large black hole.

Let us first study thermodynamics of the acoustic black hole in
the cavity based on the HUP. In an irrotational fluid, 
the propagation of sound waves is governed
by the equation of motion~\cite{unruh},
\begin{equation}
  \label{eom}
  \Box \psi = \frac{1}{\sqrt{-g}} \partial_\mu (\sqrt{-g} g^{\mu\nu}
  \partial_\nu \psi)=0,
\end{equation}
where $\psi$ is the fluctuation of the velocity potential
interpreted as a sonic wave function, and the metric is given by
\begin{equation}
  \label{g}
  g_{\mu\nu} = \frac{\rho_0}{c} \left(
    \begin{array}{cc}
      -(c^2 - v_0^2) & - v_0^i \\
      -v_0^j & \delta_{ij}
    \end{array}\right) \qquad \mathrm{with}\ i,j = 1,2,3,
\end{equation}
where $c$ is the speed of sound wave, $\rho_0$ and $v_0^i$ are the
mass density and the velocity of the mean flow, respectively,
and $v_0^2 = \delta_{ij} v_0^i
v_0^j$. Note that the velocity potential has been linearized as $\Psi =
\psi_0 + \psi$ and $\vec{v}_0 = \vec\nabla \psi_0$.
We then consider a draining bathtub fluid flow described by a
$(2+1)$-dimensional flow with a sink at the origin. If the metric is
stationary and axisymmetric according to the equation of continuity, Stokes'
theorem, and conservation of angular momentum, then $\rho_0$ is
actually a constant and $\psi_0 (r,\phi) = - A \ln (r/a) + B\phi$, where $a$,
$A$, and $B$ are arbitrary real positive constants~\cite{visser}.
So, the velocity of the mean flow in the draining vortex is given by $\vec{v}_0 = -
\hat{r} (A/r) + \hat{\phi} (B/r)$.

Dropping the position-independent prefactor in 
Eq.~(\ref{g}), the acoustic line element of the
draining bathtub is given by
\begin{equation}
  \label{line}
  ds^2 = -c^2 dt^2 + \left( dr + \frac{A}{r}dt \right)^2 + \left(r
  d\phi - \frac{B}{r}dt \right)^2,
\end{equation}
where the radii of the horizon and the ergosphere are
\begin{equation}
  \label{horizon}
  r_H = \frac{A}{c}, \qquad r_e = \frac{\sqrt{A^2 + B^2}}{c},
\end{equation}
respectively. Using the following coordinate transformation 
in the exterior region of $A/c < r < \infty$~\cite{bm,bcl},
\begin{equation}
  \label{transf}
  dt \rightarrow dt + \frac{Ar}{r^2c^2 - A^2} dr, \qquad d\phi \rightarrow
  d\phi + \frac{AB}{r(r^2c^2 - A^2)} dr,
\end{equation}
the metric~(\ref{line}) can be rewritten in the conventional form of
\begin{equation}
  \label{metric}
  ds^2 = - N^2 dt^2 + N^{-2} dr^2 + r^2 (d\phi - \Omega_0 dt)^2
\end{equation}
with
\begin{equation}
  \label{func}
  N^2(r) = 1 - \frac{A^2}{c^2r^2} = \frac{r^2 - r_H^2}{r^2}, \qquad
  \Omega_0(r) = \frac{B}{cr^2} = \Omega_H \frac{r_H^2}{r^2},
\end{equation}
where we rescaled the time coordinate by $c$ for simplicity and
$\Omega_H = B/(cr_H^2)$. Note that the metric~(\ref{metric}) looks 
similar to that of the rotating Ba\~nados-Teitelboim-Zanelli (BTZ) black
hole~\cite{btz}. However, it has a slight difference from the lapse function
$N(r)$ because it is given by $N_\mathrm{BTZ}^2 = (r^2 - r_+^2)(r^2 -
r_-^2)/(r^2l^2)$ for the BTZ case. Moreover, the acoustic black hole
is asymptotically flat while the BTZ black  hole has asymptotically
anti-de Sitter spacetime.
From now on, we will consider non-rotating case only, \textit{i.e.},
$\Omega_H=0$, 
for simplicity.

To investigate the thermodynamics of the acoustic black
holes, the entropy is assumed to satisfy the area law,
since it is independent of the detail structure of asymptotic
behavior of gravitational field or matter field. In
other words, it depends only on the geometry of the horizon.
In the three-dimensional spacetime, the
Bekenstein-Hawking entropy is given by~\cite{bcm,btz,jm}
\begin{equation}
  \label{S}
  S = \frac{4\pi r_H}{\ell_p},
\end{equation}
where the three-dimensional Plank length is chosen as $\ell_p \equiv
\hbar G/c^3$.
%
%
On the other hand, the surface gravity of the acoustic black hole is
given by $\kappa_H^2 \equiv -\left.\frac12\nabla^\mu\chi^\nu
  \nabla_\mu\chi_\nu \right|_{r=r_H} = 1/r_H^2$, where we used an
appropriate Killing field near the horizon, $\chi^\mu = (\partial_t)^\mu$~\cite{wald:book}, then the Hawking temperature becomes 
\begin{equation}
  \label{TH:HUP}
  T_H = \beta_H^{-1} = \frac{\kappa_H}{2\pi} = \frac{1}{2\pi r_H}.
\end{equation}
Heuristically, it can be induced from the HUP, 
$\Delta p = \hbar/\Delta x$. 
Putting $\Delta x = r_H$
and setting $\hbar=G=1$ for simplicity, the energy of an emitted
photon can be identified with the black hole temperature with a
``calibration factor'' $(2\pi)^{-1}$, which results in the Hawking temperature~(\ref{TH:HUP}).

Considering the acoustic black hole in a cavity with a radius $R$,
the local temperature on the boundary of the cavity is given
by~\cite{tolman}
\begin{equation}
  \label{T:HUP}
  \tilde{T} = \frac{T_H}{N(R)} = \frac{1}{2\pi r_H N(R)}.
\end{equation}
Note that the temperature of the black hole is necessarily higher than the critical
temperature $\tilde{T}_c = R/\sqrt{2}$ and there
exist both small and large black holes for a given temperature $\tilde{T}>\tilde{T}_c$.
From the first law of thermodynamics $d\tilde{E} = \tilde{T}dS$ for
a fixed $R$, we obtain the thermodynamic energy as
\begin{equation}
  \label{E:HUP}
  \tilde{E} = \frac{2}{\ell_p} \ln \left[
  \frac{r_H/\ell_p}{1+\sqrt{1-r_H^2/R^2}} \right] + \tilde{u}(R), 
\end{equation}
where $\tilde{u}(R)$ is an arbitrary integration function which
does not affect the other thermodynamic quantities.
Note that we cannot
avoid the infinite negative thermodynamic energy when
$r_H$ approaches zero.
Fortunately, this divergence can be removed by introducing minimal
length through the GUP, which will be discussed in later.
On the other hand, as for the black hole stability, it can be determined by the heat
capacity,
\begin{equation}
  \label{C:HUP}
  \tilde{C}_R = \tilde{T} \left( \frac{\partial S}{\partial
    \tilde{T}} \right)_R = \left( \frac{\partial \tilde{E}}{\partial
    \tilde{T}} \right)_R = -\frac{4\pi r_H}{\ell_p} \frac{R^2 -
  r_H^2}{R^2-2 r_H^2},
\end{equation}
which shows that the small acoustic black hole
in $r_H<R/\sqrt{2}$ is unstable and the large one
in the region of $R/\sqrt{2} < r_H < R$ is stable.
Even though this acoustic black hole is defined
in the three dimensions, the stability structure is closer to that of the
Schwarzschild black hole rather than the BTZ black hole, since the
nonrotating BTZ black hole is always stable.                      

Now, the GUP will modify the Hawking temperature and then it affects
behaviors of thermodynamic quantities. 
The HUP can be generalized into the GUP~\cite{pad,maggiore,garay,kmm,kln} based on some properties of
quantum gravity and string theory, which is actually given by 
\begin{equation}
  \Delta x \Delta p \ge \hbar \left( 1 +
      \ell^2 \frac{(\Delta p)^2}{\hbar^2} \right), \label{GUP}
\end{equation}
where it leads to the minimal length of $\Delta x_{\rm min} =
2\ell$. The cutoff $\ell$ may be chosen as a string scale in the
context of the perturbative string theory or Plank scale based on the
quantum gravity. In this GUP, the energy of the emitted photon can be identified with
the black hole temperature similar to the HUP case so that 
the momentum uncertainty is written as
\begin{equation}
  \label{eq:dp}
  \Delta p = \frac{\Delta x}{2\ell^2} \left[ 1 \pm \sqrt{ 1 -
      \frac{4\ell^2}{(\Delta x)^2}} \right].
\end{equation}
Here, we have to choose negative sign in Eq.~(\ref{eq:dp}),
since it gives the correct Hawking temperature in the HUP for the
large black hole~\cite{acs}.
Along with an appropriate calibration factor, 
the modified Hawking temperature can be written as
\begin{equation}
  \label{TH:GUP}
  T_\mathrm{GUP} = \frac{r_H}{4\pi\ell^2} \left[ 1 - \sqrt{ 1 -
      \frac{4\ell^2}{r_H^2}} \right],
\end{equation}
which is well-defined only for $r_H>2\ell$. Of course,
$T_\mathrm{GUP}$ goes to the Hawking temperature~(\ref{TH:HUP})
when the minimal length vanishes.
It is interesting to note that there exists a minimal black hole
whose radius $r_H = 2\ell$ with the finite temperature in contrast to
the Hawking temperature in the HUP.

\begin{figure}[pt]
  \includegraphics[width=0.5\textwidth]{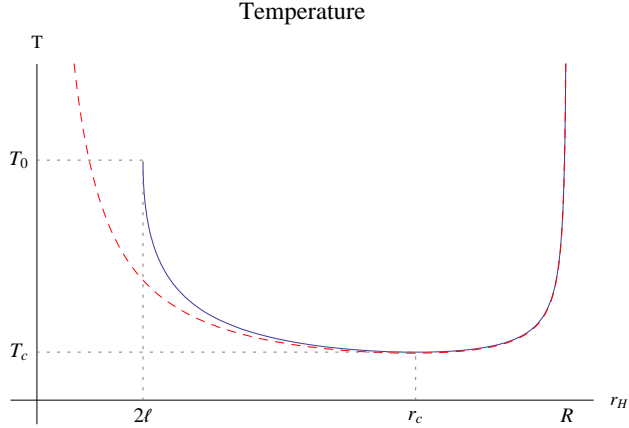}
  \caption{The dashed line and the solid line show the 
        profiles of the temperature based on the HUP and the GUP,
        respectively. For $\ell_p = 1$, $\ell=1$, and $R=10$, we have $r_c = 50/7$, $T_c =
    1/(4\pi\sqrt{6})$, and $T_0 = 5/(4\pi\sqrt{6})$.}
  \label{fig:T}
\end{figure}

\begin{figure}[pt]
  \includegraphics[width=0.5\textwidth]{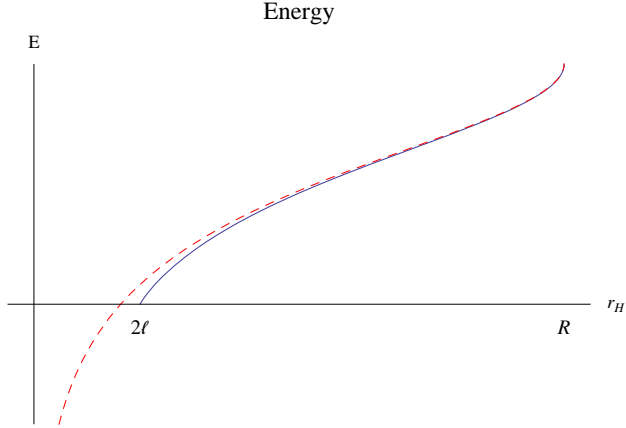}
  \caption{The dashed line and the solid line describe the 
        behaviors of thermodynamic energy based on the HUP and the GUP,
        respectively. To plot this figure, we used the same data in
        Fig.~\ref{fig:T}}
  \label{fig:E}
\end{figure}

For the cavity with a radius $R$, the local temperature becomes
\begin{equation}
  \label{T:GUP}
  T = T_\mathrm{GUP}/N(R),
\end{equation}
where $r_H < R$.
As seen from Eq.~(\ref{T:GUP}), the temperature of the black hole is always
higher than the critical temperature $T_c = T|_{r_H=r_c}$, where the
critical radius is given by $r_c = R^2/\sqrt{2(R^2-2\ell^2)}$. 
There are no black hole states for $T<T_c$ and there
can exist both small and large black holes for $T_c < T < T_0$ where
$T_0 = T|_{r_H = 2\ell}$ $ = (2\ell)^{-1}R T_c$ whereas only the
large black hole state is possible for $T > T_0$, as shown in Fig.~\ref{fig:T}.
Then, from the thermodynamic first law $dE = TdS$ for a fixed $R$ 
using the area law~(\ref{S}) and the temperature~(\ref{T:GUP}), we obtain the thermodynamic energy as
\begin{equation}
  \label{E:GUP}
  E = E_1 + u(R),
\end{equation}
with
\begin{equation}
  \label{E1:GUP}
  E_1 = - \frac{1}{\ell_p \ell^2}\left[ \int_{2\ell}^{r_H} dr\,
    \frac{\sqrt{r^2 - 4\ell^2}}{\sqrt{1-r^2/R^2}} + R^2 \left(
      \sqrt{1-\frac{r_H^2}{R^2}}- \sqrt{1-\frac{4\ell^2}{R^2}} \right)
  \right],
\end{equation}
where $E_1$ has been normalized to zero for the minimal black hole of
$r_H = 2\ell$ and $u(R)$ is an arbitrary finite function depending only on
$R$. For the case of a large black hole, the thermodynamic energy from
the GUP is the same with that of the HUP,
since the GUP effect 
is negligible for the large black hole, which gives
the matching condition as 
$u(R) = (\tilde{E} - E_1)|_{r_H=R}$ from Eqs.~(\ref{E:HUP}) and (\ref{E:GUP}). 
The GUP makes the divergent energy of the HUP finite and positive 
while the energy of the HUP is negative and divergent as the
event horizon approaches zero, which is shown in Fig.~\ref{fig:E}. 
As expected from the above matching
condition, these two thermodynamic energies are coincident for a large
black hole comparable to the cavity size.

\begin{figure}[pt]
  \includegraphics[width=0.5\textwidth]{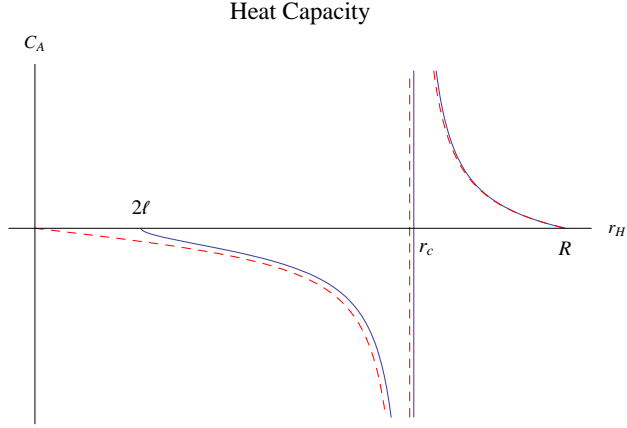}
  \caption{The crucial difference of the heat capacity from the HUP
    case comes from the end state of the black hole, since the stable minimal
    black hole exists in contrast to the conventional case shown in
    the dotted line.}
  \label{fig:C}
\end{figure}

In order to check the thermodynamic stability of the black hole, 
we should calculate the
heat capacity,  
\begin{equation}
  \label{C:GUP}
  C_R = \left(
    \frac{\partial E}{\partial T} \right)_R = - \frac{4\pi}{\ell_p}
   \frac{(R^2 - r_H^2) \sqrt{r_H^2 - 4\ell^2}}{R^2 - r_H(r_H +
     \sqrt{r_H^2 - 4\ell^2})},
\end{equation}
which is negative (unstable) for $2\ell<r_H<r_c$ and
positive (stable) for $r_c<r_H<R$ as shown in Fig.~\ref{fig:C}.
So, the small black hole less than the critical horizon is unstable while
the large black hole is stable. 
It is interesting to note that the heat capacity goes to
zero as the horizon approaches the minimal black hole so that 
it is no longer unstable and may remain as a remnant.
Actually, this acoustic black hole is different from the Schwarzschild
black hole in the GUP, since its heat capacity is negative at
the minimal size of the black hole, which is unstable~\cite{ksy}. 

On the other hand, there is another way to discuss the thermodynamic stability of the black hole.
The off-shell free energy in the cavity is
given by 
\begin{equation}
  \label{F}
  F = E - TS,
\end{equation}
where $E$ and $S$ are given by Eqs.~(\ref{S}) and (\ref{E:GUP}),
while $T$ is an arbitrary temperature. 
The free energy of the minimal black hole 
has the negative free energy for all temperatures, since 
the energy becomes zero and the entropy remains finite as $r_H\to2\ell$.
\begin{figure}[pt]
  \includegraphics[width=0.5\textwidth]{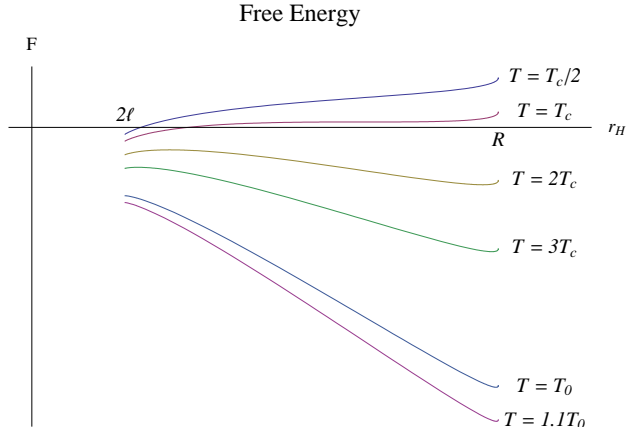}
  \caption{The free energy has two extrema for $T_c<T<T_0$ and one
    minimum for $T>T_0$ while there are no extrema for $T<T_c$.}
  \label{fig:F}
\end{figure}
Then, the extrema of the off-shell free energy can be obtained from
\begin{equation}
  \label{F:extrema}
  \left( \frac{\partial F}{\partial r_H} \right)_{R,T} = 0,
\end{equation}
which yields nothing but the temperature relation~(\ref{T:GUP}). 
Since there are no black hole states for $T<T_c$,
the free energy has no extrema. Note that it has two extrema 
corresponding to the unstable small black hole (maximum) and the stable large
black hole (minimum) for $T_c < T < T_0$ while there exists only
a stable large black hole state for $T>T_0$ as seen in
Fig.~\ref{fig:F}. All these features are compatible with 
the previous discussion in Fig.~\ref{fig:T}.  

We have studied the thermodynamics of
the acoustic black hole in the cavity based on the HUP and the GUP. 
In the former case,
the energy may become negative and eventually diverges as the 
black hole evaporates completely 
and this divergence cannot be removed by any normalizations.
However, with the help of the minimal length in the GUP regime, 
the divergence of the thermodynamic energy does not appear anymore so
that we can normalize it to zero for the minimal black hole.
Since there is a minimized horizon, it implies that there exists another critical 
temperature $T_0$ as in the Schwarzschild black
hole~\cite{acs,ch,mkp,ksy} 
such that both small and large black hole states are possible within $T_c < T
< T_0$ and there exists only a large black hole for $T>T_0$. 
Note that the small black hole is unstable while the large one is
stable, because the heat capacity is negative for $r_H<r_c$ and positive for
$r_H>r_c$. The unstable small black hole may shrink 
until its size reaches the minimal length $r_H=2\ell$,
then the acoustic black hole is no longer unstable and may remain as a remnant.

As a final comment, the Hawking temperature generically depends on the charges such as the mass, the electric charge, and the angular momentum. The metric describing the black hole geometry contains the charges and then the Hawking temperature reflects its hairs through the Euclidean time formulation. However, in this GUP regime, we can not obtain the Hawking temperature directly from the metric so that the GUP improved Hawking temperature is still unknown for the rotating geometry. In our formulation, the GUP effect does not appear in the metric contents, which means that it is difficult to understand how to consider the back reaction of the geometry. We hope this problem will be solved elsewhere.

\acknowledgments
This work was supported by the Sogang Research Grant, 20071063 (2007)
and the Science Research Center Program of the Korea Science and
Engineering Foundation through the Center for Quantum Spacetime
(CQUeST) of Sogang University with grant number R11-2005-021.


\end{document}